\documentclass[12pt,reqno]{article}
\usepackage{epsfig}
\usepackage{amsmath}
\usepackage{amssymb}
\usepackage{amsthm}
\allowdisplaybreaks[4]
\numberwithin{equation}{section}
\date{}
\begin{document}

\renewcommand{\theequation}{\arabic{section}.\arabic{equation}}
\newcommand{\aknsh}{AKNS hierarchy\ }
\newcommand{\lo}{Lax operator\ }  
\newcommand{\gt}{gauge transformation\ }
\newcommand{\gts}{gauge transformations\ }
\newcommand{\gto}{gauge transformation operator\ }
\newcommand{\gtos}{gauge transformation operators\ }
\newcommand{\cgts}{chain of gauge transformations\ }
\newcommand{\hy}{hierarchy\ }
\newcommand{\ckph}{cKP hierarchy\ }
\newcommand{\tf}{tau-function}
\newcommand{\La}[1]{L^{({#1})}}
\newcommand{\Laa}[1]{L^{^*{({#1})}}}
\newcommand{\B}[2]{B^{({#1})}_{#2}}
\newcommand{\Ba}[2]{B^{^*({#1})}_{#2}}
\newcommand{\T}[3]{T^{({#1})}_{{#2}}({#3})}
\newcommand{\fun}[3]{{#1}^{({#2})}_{{#3}}}
\newcommand{\tu}[1]{\tau^{(#1)}}
\newtheorem{theorem}{Theorem}
\newtheorem{lemma}{Lemma}
\newtheorem{corollary}{Corollary}
\newtheorem{remark}{Remark}
\title{\Huge Solutions  of the (2+1)-dimensional KP, SK and KK equations generated  by gauge transformations from non-zero
seeds } \maketitle
\author{\large \begin{center}
         $\begin{array}{l}
             \text {\sl Jingsong He , Xiaodong Li}\\
             \text {\small Department of  Mathematics }\\
             \text {\small University of Science and Technology of China} \\
             \text {\small Hefei, 230026 Anhui}\\
             \text {\small P.R. China }
          \end{array}$
      \end{center}
      }

\vspace{1cm}
\abstract{} \noindent By using gauge transformations, we manage to
obtain new solutions of ($2+1$)-dimensional
Kadomtsev-Petviashvili(KP), Kaup-Kuperschmidt(KK) and
Sawada-Kotera(SK) equations from non-zero seeds. For each of the
preceding equations, a Galilean type transformation between these
solutions $u_2$ and the previously known solutions $u_2^{\prime}$
generated from zero seed is given.  We present several explicit
formulas of the single-soliton solutions for $u_2$ and
$u_2^{\prime}$, and further point out  the two main differences of
them under the same value of parameters, i.e., height and location
of peak line, which are demonstrated visibly in  three figures.
\section{Introduction}
In the 1980s, Sato and his colleagues brought us the famous Sato
theory \cite{tau,Jimbo1}.  Since then, the pseudo-differential
operator has been playing an important role in the research of the
Kadomtsev-Petviashvili(KP) hierarchy \cite{d}, which can yield many
important nonlinear partial differential equations, such as the
generalized nonlinear Schr\"{o}dinger equation, the KdV equation,
the Sine-Gordon equation and the famous KP equation. To be
self-consistent, we would like to give a brief review of the KP
hierarchy \cite{tau,Jimbo1,d,Jimbo2}.

Let
\begin{equation}\label{laxoperatorkP}
L=\partial+u_2\partial^{-1}+u_3\partial^{-2}+\cdots ,
\end{equation}
be a pseudo-differential operator($\Psi$DO), here $\{u_{i}\},
u_{i}=u_{i}(t_{1},t_{2},t_{3},\ldots)$ serve as generators of a
differential algebra $\mathcal{A}$. The corresponding generalized
Lax equations are defined as
\begin{equation}
\label{lax} \frac{\partial L}{\partial t_{n}}=[B_{n},L],\quad
n=1,2,3,\ldots,
\end{equation}
which give rise to infinite number of partial differential equations
of the KP hierarchy, $B_{n}$ is defined as $B_{n}=[L^{n}]_{+}$. It
can be easily showed that eq.(\ref{lax}) is equivalent to the
so-called Zakharov-Shabat(ZS) equation \cite{zs}
\begin{equation}
\label{zs} \frac{\partial B_{m}}{\partial t_{n}} - \frac{\partial
B_{n}}{\partial t_{m}}+[B_{m},B_{n}]=0  ,\quad  (m,n=2,3,\ldots).
\end{equation}
The eigenfunction $\phi$ and conjugate eigenfunction $\psi$
corresponding to $L$ are defined by
\begin{eqnarray}
\label{ei} \frac{\partial \phi}{\partial t_{n}} & = & B_{n} \phi,  \\
\label{cei} \frac{\partial \psi}{\partial t_{n}} & = & -B_{n}^{*}
\psi.
\end{eqnarray}
The first non-trivial example is the KP equation given by the
$t_2$-flow and $t_3$-flow of the KP hierarchy
\begin{equation}
\label{kp} (4u_{t}-12uu_{x}-u_{xxx})_{x}-3u_{yy}=0,
\end{equation}
in which $x=t_1$, $y=t_2$ and $t=t_3$.

Suppose  $L$ given by eq.(\ref{laxoperatorkP}) and  $L^{*}$ defined
by
\begin{equation*}
L^{*}=-\partial + \sum_{i=1}^{\infty} (-1)^{i}\partial^{-i}u_{i+1}.
\end{equation*}
If $L$ satisfies $L^{*}+L=0$ , then $L$ is called the Lax operator
of the CKP hierarchy \cite{Jimbo2,dkjm3}, and the corresponding flow
equations of the CKP hierarchy are described by
\begin{equation}\label{BKPflows}
\frac{\partial L}{\partial t_{n}}=[B_{n},L],\quad n=1,3,5,\cdots.
\end{equation}
The first non-trivial example is the CKP equation \cite{Jimbo2,l2}
\begin{equation}\label{CKPequation}
 u_{t} =
\frac{5}{9}\left(
\partial_x^{-1}u_{yy}+3u_x\partial_{x}^{-1}u_y-\frac{1}{5}u_{xxxxx}-
3uu_{xxx}-\frac{15}{2}u_xu_{xx}-9u^2u_{x}+u_{xxy}+3uu_{y}
    \right)
\end{equation}
which is generated by $t_3$-flow and $t_5$-flow and also called the
($2+1$)-dimensional Kaup-Kuperschmidt(KK) equation \cite{BV}. Here
$x=t_1$, $y=t_3$ and $t=t_5$. Moreover, $L$ is called the Lax
operator of the BKP hierarchy \cite{Jimbo1,dkjm2} if it satisfies
$L^{*}=-\partial L
\partial^{-1}$, and the flow equations of the BKP hierarchy
associated with it are also described by eq.(\ref{BKPflows}). The
first non-trivial example is the BKP equation \cite{l,IR}
\begin{equation}\label{BKPequation}
    u_t = \frac{5}{9}\left( \partial_x^{-1}u_{yy}+3u_x\partial_{x}^{-1}u_y-\frac{1}{5}u_{xxxxx}-3uu_{xxx}-3u_xu_{xx}-9u^2u_{x}+u_{xxy}+3uu_{y}
    \right),
\end{equation}
which is generated by $t_3$-flow and $t_5$-flow and also called the
($2+1$)-dimensional Sawada-Kotera(SK) equation \cite{BV}. Here
$x=t_1$, $y=t_3$ and $t=t_5$.

If we find a set of functions ${u_{2},u_{3},\ldots}$ which makes the
corresponding pseudo-differential operator $L$ satisfies
eq.(\ref{zs}), then we have a solution of the KP hierarchy. It's a
well-known result that this set of solutions can be generated from
one single function $\tau(x)$ as the following way
\begin{eqnarray}
\label{u2} u_{2}&=&\frac{\partial^2}{\partial t_{1}^2} \log{\tau},
\\ \label{u3}
 u_{3}&=&\frac{1}{2} [\frac{\partial^2}{\partial t_{1} \partial t_{2}} -
\frac{\partial^3}{\partial t_{1}^{3}}] \log{\tau},
\end{eqnarray}
\begin{center}
$\vdots$
\end{center}

During the last two decades, in order to solve the KP hierarchy, the
gauge transformation was formally introduced in reference
\cite{csy}. The basic idea behind gauge transformation is to find  a
transformation for the initial Lax operator $L^{(0)}$ of the KP
hierarchy after which the new operator $L^{(1)}$ and $B_n^{(1)}$
still satisfies Lax equation eq.(\ref{lax}) and eq.(\ref{zs})
respectively. Here
\begin{equation}
L^{(1)}=T \circ L^{(0)} \circ T^{-1}, \qquad
B_n^{(1)}=(L^{(1)})_+^n,
\end{equation}
$T$ is a suitable pseudo-differential operator. There exist two
kinds of gauge transformation operators \cite{csy}
\begin{eqnarray}
T_{D}(\phi^{(0)})&=&\phi^{(0)} \, \partial \, (\phi^{(0)})^{-1},\\
T_{I}(\psi^{(0)})&=&(\psi^{(0)})^{-1} \,
\partial^{-1} \, \psi^{(0)},
\end{eqnarray}
in which $\phi^{(0)}$, $\psi^{(0)}$ are eigenfunction and conjugate
eigenfunction of $L^{(0)}$ respectively and they are also called the
generating functions of the gauge transformation. $T_{D}$ is called
differential type of gauge transformation, $T_{I}$ is called
integral type of gauge transformation. After one gauge
transformation $T_{D}$, the new $\tau$-function
\begin{equation}
\label{tau1} \tau^{(1)}=\phi^{(0)} \tau^{(0)},
\end{equation}
is transformed from an initial $\tau$-function $\tau^{(0)}$
associated with the initial Lax operator $L^{(0)}$.
 A similar result can be formulated for the case of $T_{I}$
\begin{equation}
\label{tau2} \tau^{(1)}=\psi^{(0)} \tau^{(0)}.
\end{equation}
With the help of  formulas eq.(\ref{u2}), eq.(\ref{u3}),
eq.(\ref{tau1}) and eq.(\ref{tau2}), we can obtain new solutions
\{$u_i^{(1)}$\} from the known  seed solutions \{$u_i^{(0)}$\} in
the $L^{(0)}$. For example, $u_2^{(1)}=u_2^{(0)}+(\log
\phi^{(0)})_{xx}$ by the gauge transformation in eq.(\ref{tau1}). By
a successive application of gauge transformations, the determinant
representation of $\tau^{(n+k)}$ is given in \cite{hlc2} and further
more $u_2^{(n+k)}$ can be deduced by using eq.(\ref{u2}).

In the last decade, the method of gauge transformation has been
developed by several researchers. The original form of this
transformation proposed in reference \cite{csy}  cannot be applied
directly to the  sub-hierarchies of the KP hierarchy. So in
\cite{Nimmo,Amei,Jhep}, an improvement was made which makes it
applicable to the BKP and CKP hierarchies, and in
\cite{Oevel,cst,Aratyn,Willox,hlc,hwc} another improvement was made
so that the gauge transformation can be used on the constrained KP
hierarchy.  Besides gauge transformation, some other methods have
been used to solve the KP, BKP, CKP equations. In \cite{solvekp1},
Hirota method was considered on the KP equation. Darboux
transformation was applied on this equation in Chapter 3 of
\cite{solvekp2}. N-soliton solutions of the BKP equation was
obtained through Hirota method in \cite{hirota1,hirota2}, lump
solutions was obtained through this method in \cite{LumpBKP}, the
same method was applied to the ($2+1$)-dimensional KK equations in
\cite{hu} and 3-soliton solutions were obtained explicitly. Darboux
transformation was applied to ($2+1$)-dimensional KK, SK equations
in \cite{SN}. In \cite{dl}, $\bar{\partial}$- dressing method was
used on the ($2+1$)-dimensional KK, SK equations and line solitons
and line rational lumps were obtained.  It is easy to recognize that
all these known solutions are corresponding to the solutions given
by gauge transformation from zero seed. However, solving the soliton
equations starting from a "non-zero seed" has not attracted enough
attention. There are very few works on the KPI and  KP II equations
with a non-decay initial background\cite{fokas1,fokas2,ablowitz1} by
dressing method and classical inverse scattering method. On the other hand,
gauge transformation from non-zero seeds was not considered before to our knowledge.
One possible reason is that in
the case of the KdV equation, solutions obtained by gauge
transformation from zero seed can be transformed to those solutions
from non-zero seeds by a Galilean transformation \cite{tian}. So
far, we have not seen any similar discussions on solutions of
($2+1$)-dimensional KP, KK, SK equations. Therefore in this paper,
we solve these equations by gauge transformation from non-zero seeds
and manage to find out the relations between new solutions and those
from zero seed.

The organization of this paper is as follows. In section two we
consider the KP equation. In section three and section four, we
discuss (2+1)-dimensional KK and SK equations respectively. Section
five is devoted to the conclusions and discussions. The notations we
use in this paper is the same as in \cite{cst}.

\section{Successive Gauge transformation for KP equation}
It's a natural thought to consider successive application of gauge
transformation for KP hierarchy. In \cite{csy,hlc2}, a very useful
theorem was introduced about the result after successive gauge
transformations.
  \begin{lemma}[\cite{csy,hlc2}]\label{lemma1}
    After $n$ times $T_{D}$ and $k$ times $T_{I}$ transformations $(n \ge k)$ , we have :
    \begin{align}
      \nonumber \tau^{(k+n)} &= \psi_{k}^{(k-1+n)}\cdot \psi_{k-1}^{(k-2+n)} \cdots \cdot \psi_{1}^{(n)} \cdot \tau^{(n)} \\
      &={\rm IW}_{k,n}(\psi_{k}^{(0)},\psi_{k-1}^{(0)},\cdots,\psi_{1}^{(0)};\phi_{1}^{(0)},\phi_{2}^{(0)},\cdots,\phi_{n}^{(0)})\cdot\tau^{(0)},
    \end{align}
    in which ${\rm IW}_{k,n}(\psi_{k}^{(0)} , \psi_{k-1}^{(0)} , \cdots ,\psi_{1}^{(0)};\phi_{1}^{(0)}
    ,\phi_{2}^{(0)},\cdots,\phi_{n}^{(0)})$ stands for
    \begin{equation*}
    {\rm IW}_{k,n}=\begin{vmatrix}
      \int \phi_{1}^{(0)}\cdot\psi_{k}^{(0)} & \int \phi_{2}^{(0)}\cdot\psi_{k}^{(0)} & \cdots & \int \phi_{n}^{(0)}\cdot\psi_{k}^{(0)} \\
      \int \phi_{1}^{(0)}\cdot\psi_{k-1}^{(0)} & \int \phi_{2}^{(0)}\cdot\psi_{k-1}^{(0)} & \cdots & \int \phi_{n}^{(0)}\cdot\psi_{k-1}^{(0)} \\
      \vdots & \vdots & \cdots & \vdots  \\
      \int \phi_{1}^{(0)}\cdot\psi_{1}^{(0)} & \int \phi_{2}^{(0)}\cdot\psi_{1}^{(0)} & \cdots & \int \phi_{n}^{(0)}\cdot\psi_{1}^{(0)} \\
      \phi_{1}^{(0)} & \phi_{2}^{(0)} & \cdots & \phi_{n}^{(0)}  \\
      \phi_{1,x}^{(0)} & \phi_{2,x}^{(0)} & \cdots & \phi_{n,x}^{(0)} \\
      \vdots & \vdots & \cdots & \vdots  \\
      (\phi_{1}^{(0)})^{(n-k-1)} & (\phi_{2}^{(0)})^{(n-k-1)} & \cdots & (\phi_{n}^{(0)})^{(n-k-1)}
    \end{vmatrix}
    \end{equation*}
    $\phi_{i}^{(0)}$ and $\psi_{i}^{(0)}$ are solutions of
    eq.(\ref{ei}) and eq.(\ref{cei}) associated with the initial value
    $\tau^{(0)}$, further we have
    \begin{equation}\label{kpgtnplusk}
    u_{2}^{(k+n)} =
    (\log{{\rm IW}_{k,n}})_{x,x}+u_{2}^{(0)}.
    \end{equation}
  \end{lemma}

By using the above theorem, we now start to construct the new
solutions of the KP equation in eq.(\ref{kp}) from non-zero seeds.
To the end, we choose the initial Lax operator of the KP hierarchy
to be
\begin{equation*}
\label{L1} L^{(0)}=\partial + \partial^{-1} + \partial^{-2} +
\partial^{-3} + \cdots,
\end{equation*}
such that all $u_{i}^{(0)}=1$ and  then the seed solution of the KP
equation  is $u^{(0)}=u_2^{(0)}=1$. We know that the KP equation is
generated by $t_2$-flow and $t_3$-flow of the KP hierarchy, so the
generating functions {$\phi_{i}^{(0)}$} and {$\psi_{i}^{(0)}$} for
the gauge transformation satisfy
\begin{eqnarray}
\label{t11}&
\begin{cases}
\phi_{i,t_2}^{(0)} & =  B^{(0)}_{2} \phi_{i}^{(0)} = (\partial^{2} + 2)\phi_{i}^{(0)},\quad  B^{(0)}_{2}= (L^{(0)})^2_{+} \\
\phi_{i,t_3}^{(0)} & =  B^{(0)}_{3} \phi_{i}^{(0)} = (\partial^{3} +
3\partial +3)\phi_{i}^{(0)},  \quad  B^{(0)}_{3}= (L^{(0)})^3_{+}
\end{cases}&\\&\label{t12}
\begin{cases}
\psi_{i,t_2}^{(0)} & =  -(B_{2}^{(0)})^{*} \psi_{i}^{(0)} = -(\partial^{2} + 2)\psi_{i}^{(0)}, \\
\psi_{i,t_3}^{(0)} & =  -(B_{3}^{(0)})^{*} \psi_{i}^{(0)} =
(\partial^{3} + 3\partial -3)\psi_{i}^{(0)}.
\end{cases}&
\end{eqnarray}

\begin{lemma}\label{lemma2}
The solutions of eq.(\ref{t11}), eq.(\ref{t12}) are in form of
\begin{eqnarray}
\label{kpphi} \phi_{i}^{(0)} &=& \sum_{j=1}^{n} k_{j}
e^{\frac{\beta_{j}-3}{\alpha_{j}+1} x +\alpha_{j} y + \beta_{j} t}
,\quad \beta_{j}=\beta_{j}(\alpha_{j}),\\
\label{kppsi}\psi_{i}^{(0)} &=& \sum_{j=1}^{m} \widetilde{k_{j}}
e^{\frac{\widetilde{\beta_{j}}+3}{-\widetilde{\alpha_{j}}+1} x
+\widetilde{\alpha_{j}} y + \widetilde{\beta_{j}} t} ,\quad
\widetilde{\beta_{j}}=\widetilde{\beta_{j}}(\widetilde{\alpha_{j}}).
\end{eqnarray}
Here $\alpha_{j}$ , $\beta_{j}$ ,$\widetilde{\alpha_{j}}$ ,
$\widetilde{\beta_{j}}$ should satisfy the following relations
\begin{eqnarray}
\label{constrain1}
(\beta_{j}-3)^2 &=& (\alpha_{j} +1)^2 (\alpha_{j} -2), \\
(\widetilde{\beta_{j}}+3)^2 &=& (-\widetilde{\alpha_{j}} +1)^2
(-\widetilde{\alpha_{j}} -2).
\end{eqnarray}
\end{lemma}
\begin{proof}
We assume the solutions of eq.(\ref{t11}) have the form
$\widehat{\phi} = X(x)\,Y(y)\,T(t)$, then eq.(\ref{t11}) is
equivalent to
\begin{eqnarray}
  \begin{cases}
    \frac{Y_y}{Y} & = \frac{X_{xx}}{X} + 2, \\
    \frac{T_t}{T} & = \frac{X_{xxx}}{X} + 3\,\frac{X_x}{X} + 3.
  \end{cases}
\end{eqnarray}
Let
\begin{equation}\label{YT1}
    \frac{Y_y}{Y} = \alpha, \qquad
    \frac{T_t}{T} = \beta,
\end{equation}
where $\alpha$ and $\beta$ are constants, we have
\begin{eqnarray}
  \begin{cases}
    (\alpha - 2)\,X & = X_{xx}, \\
    (\beta - 3)\,X & =  X_{xxx} + 3\,X_{x},
  \end{cases}
\end{eqnarray}
which can be reduced to
\begin{eqnarray}
  \begin{cases}
    X_{x} & = \frac{(\alpha+1)\,(\alpha-2)}{\beta-3}\,X, \\
    X_{x} & = \frac{\beta-3}{\alpha+1}\,X.
  \end{cases}
\end{eqnarray}
Under the consistency condition $(\beta-3)^2 =
(\alpha+1)^2\,(\alpha-2)$ we can obtain
\begin{equation}
\label{X} X(x) = c_1\, e^{\frac{\beta-3}{\alpha+1}\,x}.
\end{equation}
From eq.(\ref{YT1}), we have
\begin{equation*}
  Y(y)  =  c_2\, e^{\alpha y}, \qquad
  T(t)  =  c_3\, e^{\beta t},
\end{equation*}
which infer the solutions of eq.(\ref{t11})
\begin{equation}\label{lemma21}
  \widehat{\phi} = k\,e^{\frac{\beta-3}{\alpha+1}\,x+\alpha y+\beta
  t},
\end{equation}
with the help of (\ref{X}),
where $k = c_1 \, c_2 \, c_3$.  By linear superposition, the linear
combination of $\widehat{\phi}$ in eq.(\ref{lemma21})
with respect to  different $\alpha$ and $\beta$ is still a solution of eq.(\ref{t11}), that
is
\begin{equation}
  \phi_{i}^{(0)} = \sum_{j=1}^{n} k_{j} \, \widehat{\phi_j} = \sum_{j=1}^{n}
  k_{j} \,
e^{\frac{\beta_{j}-3}{\alpha_{j}+1} x +\alpha_{j} y + \beta_{j} t}
\end{equation}
A similar procedure can be applied to $\psi_{i}^{(0)}$ which yields eq.(\ref{kppsi}).
\end{proof}
Having these results, it's sufficient to perform gauge
transformation on $L^{(0)}$. But according to lemma \ref{lemma1},
the transformed $\tau$-function may not be satisfactory, since it
may vanish on some point. To rule out this situation, we need the
following theorem.
  \begin{theorem}\label{theorem1}
  Let the generating functions of n-steps $T_{D}$ be  $\phi_{m}^{(0)}\, ( m=1,2,\cdots,n)$ in eq.(\ref{kpphi}) and rewritten as $\phi_m^{(0)} = \sum_{i=1}^{p_m} k_{m,i}\exp^{a_{m,i}x+\alpha_{m,i}y+\beta_{m,i}t}$ for simplicity, then the new $\tau$-function
  \begin{equation}
  \label{kpp}
    \tau^{(n)} = {\rm IW}_{0,n} \cdot \tau^{(0)} =
    {\rm W}_{n}(\phi_{1}^{(0)},\phi_{2}^{(0)},\cdots,\phi_{n}^{(0)})
    \cdot \tau^{(0)},
  \end{equation}
  and ${\rm W}_{n}(\phi_{1}^{(0)},\phi_{2}^{(0)},\cdots,\phi_{n}^{(0)})>0$
  if $k_{m,i}>0$ , $a_{m,i}<a_{m^{'},j}$ for all $m<m^{'}$ and $\forall \, i,j$.
  The transformed solution $u_2^{(n)}$ of {\rm KP} equation  is
  \begin{equation}\label{kpu2}
    u_2^{(n)} = 1 + \left(\log\left({\rm W}_{n}(\phi_{1}^{(0)},\phi_{2}^{(0)},\cdots,\phi_{n}^{(0)})\right) \right)_{xx}
  \end{equation}
  \end{theorem}
  \begin{proof}
    First, ${\rm W}_{n}$ takes the following form
    \begin{equation*}
    {\rm W}_{n} =
        \begin{vmatrix}
            \phi_{1}^{(0)} & \phi_{2}^{(0)} & \cdots & \phi_{n}^{(0)} \\
            \frac{\partial}{\partial x} \phi_{1}^{(0)} & \frac{\partial}{\partial x} \phi_{2}^{(0)} & \cdots & \frac{\partial}{\partial x} \phi_{n}^{(0)} \\
            \vdots & \vdots & \cdots & \vdots \\
            \frac{\partial^{n-1}}{\partial x^{n-1}} \phi_{1}^{(0)} & \frac{\partial^{n-1}}{\partial x^{n-1}} \phi_{2}^{(0)} & \cdots & \frac{\partial^{n-1}}{\partial x^{n-1}} \phi_{n}^{(0)}
        \end{vmatrix}_{n \times n}
    \end{equation*}
  then we expand the determinant with respect to columns using the equation
  \begin{equation*}
        \phi_{m}^{(0)}=\sum_{i=1}^{p_m} k_{m,i}\,e^{a_{m,i}\,x+\alpha_{m,i}\,y+\beta_{m,i}\,t},\quad m=1 \ldots n.
  \end{equation*}
  then we have:
  \begin{equation}
  \label{iw0n}
    {\rm W}_{n} = \sum_{1 \leq i_q \leq p_q , \, q=1\ldots n} \Pi_{j=1}^{n}
    k_{j,i_j}\,e^{a_{j,i_j}x+\alpha_{j,i_j}y+\beta_{j,i_j}t}\,
    \begin{vmatrix}
        1 & 1 & \cdots & 1 \\
        a_{1,i_1} & a_{2,i_2} & \cdots & a_{n,i_n} \\
        \vdots & \vdots & \cdots & \vdots \\
        a_{1,i_1}^{n-1} & a_{2,i_2}^{n-1} & \cdots & a_{n,i_n}^{n-1}
    \end{vmatrix}
  \end{equation}
  Notice the Vendermonde determinants in the above equation. Since $k_{m,i}>0$, the coefficients of these Vendermonde determinants are positive. Using $a_{m,i}<a_{m^{'},j}$ for all $m<m^{'}$ and $\forall i,j$, it's easy to prove that all Vendermonde determinants in the above equation are positive, so ${\rm W}_{n}>0$.
  Using eq.(\ref{kpp}), eq.(\ref{kpgtnplusk}) and $u_2^{(0)}=1$, we can obtain eq.(\ref{kpu2}).
  \end{proof}

Next we give single-soliton solutions of the KP equation from a zero
seed and a non-zero seed respectively. Notations with prime are
corresponding to the results of gauge transformation from a zero
seed. The generating functions are
\begin{eqnarray}
\left(\phi_1^{(0)} \right)^{'} &=& k^{'} \, e^{\xi_1^{'}}
             +k^{'} \, e^{\xi_2^{'}},\\
  \phi_1^{(0)} &=& k \, e^{\xi_1}
             +k \, e^{\xi_2},
\end{eqnarray}
where
\begin{eqnarray}
\xi_1^{'} &=& {\frac {{ \beta_{1}^{'}}}{{ \alpha_{1}^{'}}}}\,x+{{
\alpha_{1}^{'}}}\,y+{{ \beta_{1}^{'}}}\,t, \\
\xi_2^{'}&=&{\frac {{ \beta_{2}^{'}}}{{ \alpha_{2}^{'}}}\, x}+{{
\alpha_{2}^{'}}}\,y+{{ \beta_{2}^{'}}}\,t, \\
\xi_1 &=& {\frac { ( { \beta_{1}}-3 ) }{{
\alpha_{1}}+1}} \,x+{ \alpha_{1}}\,y+{
 \beta_{1}}\,t, \\
 \xi_2 &=& {\frac { ( { \beta_{2}}-3 ) }{{
\alpha_{2}}+1}} \,x+{ \alpha_{2}}\,y+{
 \beta_{2}}\,t,
\end{eqnarray}
and $(\alpha_i^{'})^{3}=(\beta_i^{'})^{2}$, $(\beta_i-3)^2 =
(\alpha_i+1)^2\,(\alpha_i-2)$, $i=1,2$. The two single-solitons  of the
KP equation can be written as
\begin{eqnarray}\label{kp0seed}
  (u_2^{(1)})^{'}  & = & \frac{1}{4} \, (\frac{\beta_1^{'}}{\alpha_1^{'}} -
    \frac{\beta_2^{'}}{\alpha_2^{'}})^2 \, {\rm sech}^2 (\frac{\xi_2^{'}-\xi_1^{'}}{2}),\\ \label{kpn0seed}
  u_2^{(1)} & = & 1 + \frac{1}{4} \, (\frac{\beta_1-3}{\alpha_1+1} -
    \frac{\beta_2-3}{\alpha_2+1})^2 \,
    {\rm sech}^2 (\frac{\xi_1-\xi_2}{2}).
\end{eqnarray}
There are two differences  between $u_2$ and $u_2^{\prime }$ under
the same parameters  $\alpha$: 1) the height  of solitons, 2) the
location of the peak line of the solitons, which are demonstrated
visibly in figure 1.
 In figure 2,
we demonstrate the solution obtained by a two-step gauge
transformation by using eq.(\ref{kpu2}) and
\begin{eqnarray}
  \phi_1^{(0)} & = & e^{2 \, y + 3 \, t} + e^{x + 3 \, y + 7 \, t}, \\
  \phi_2^{(0)} & = & e^{\sqrt{2}\,x + 4 \, y + (3+5\sqrt{2}) \, t} +
  e^{\sqrt{6} \, x + 8 \, y + (3+9\sqrt{6}) \, t}.
\end{eqnarray}
\begin{corollary}\label{corkptrans}
There exists a Galilean type transformation
 \begin{eqnarray}\label{kptrans}
   u_2^{\prime} & \longmapsto & u_2(x,y,t)= 1 + u_2^{\prime}(x+3\,t, \ y,\ t).
  \end{eqnarray}
  between $u_2^{'}$ in eq.(\ref{kp0seed}) and $u_2$ in eq.(\ref{kpn0seed}).
\end{corollary}
\noindent Obviously, this result is consistent with the Galilean
transformation \cite{tian} of the KdV equation by a dimensional reduction.

\section{Gauge transformation for (2+1)-dimensional KK equation}

Gauge transformation of the CKP hierarchy is somewhat different from
that of the KP hierarchy, because a transformed Lax operator
$L^{(1)}$  by one-step gauge transformation has to satisfy
$(L^{(1)})^{*}+L^{(1)}=0$.  To meet this requirement, we introduce
the following lemma.
\begin{lemma}[\cite{Jhep}]\label{lemma3}
\begin{enumerate}
    \item The appropriate gauge transformation $T_{n+k}$ is given by $n=k$ and generating functions $\psi_{i}^{(0)}=\phi_{i}^{(0)}$ for $i=1,2,\cdots,n$.
    \item The $\tau$-function $\tau_{{\rm CKP}}^{(n+n)}$ of the {\rm CKP} hierarchy has the form
      \begin{eqnarray}\nonumber
        \tau_{\text{\rm CKP}}^{(n+n)} &=& {\rm IW}_{n,n}(\phi_{n}^{(0)},\phi_{n-1}^{(0)},\cdots,\phi_{1}^{(0)};\phi_{1}^{(0)},\phi_{2}^{(0)},\cdots,\phi_{n}^{(0)}) \cdot \tau_{\text{\rm CKP}}^{(0)}\\
        &=&
        \begin{vmatrix}
          \int \phi_{n}^{(0)}\cdot\phi_{1}^{(0)} & \cdots & \int \phi_{n}^{(0)}\cdot\phi_{n}^{(0)} \\
          \vdots & \cdots & \vdots \\
          \int \phi_{1}^{(0)}\cdot\phi_{1}^{(0)} & \cdots & \int \phi_{1}^{(0)}\cdot\phi_{n}^{(0)}
        \end{vmatrix}
        \cdot
        \tau_{\text{\rm CKP}}^{(0)}.
      \end{eqnarray}
    \end{enumerate}
    and further we have
    \begin{equation}\label{CKPgtnplusn}
    u_{2}^{(n+n)} =u_{2}^{(0)} +
    (\log{{\rm IW}_{n,n}})_{xx}.
    \end{equation}
\end{lemma}

To solve the (2+1)-dimensional KK equation from non-zero seed
solution, we choose a initial Lax operator $L^{(0)}$ of the CKP
hierarchy to be
\begin{equation*}
L^{(0)}=\partial+\partial^{-1}+\partial^{-3}+\partial^{-5}+\cdots.
\end{equation*}
Since the (2+1)-dimensional KK equation is generated by $t_3$-flow
and $t_5$-flow of the CKP hierarchy, we solve
\begin{eqnarray}
\label{t31}
\begin{cases}
\phi_{i,t_3}^{(0)} & =  B_{3}^{(0)} \phi_{i}^{(0)} = (\partial^{3} + 3\partial)\,\phi_{i}^{(0)}, \quad B_{3}^{(0)}
=(L^{(0)})^3_{+} , \\
\phi_{i,t_5}^{(0)} & =  B_{5}^{(0)} \phi_{i}^{(0)} = (\partial^{5} +
5\partial^{3} + 15\partial)\,\phi_{i}^{(0)},  \quad B_{5}^{(0)}
=(L^{(0)})^5_{+},
\end{cases}
\end{eqnarray}
in order to obtain the eigenfunctions.
\begin{lemma}\label{lemma4}
The solutions of eq.(\ref{t31}) are
\begin{equation}
\label{kkphi} \phi_{i}^{(0)}=\sum_{j=1}^{n} k_j \,
e^{\frac{\alpha_{j}^{3}-18\alpha_{j}+9\beta_{j}}{\alpha_{j}^{2}+\alpha_{j}\beta_{j}+81}
x +\alpha_{j} y + \beta_{j} t}, \quad
\beta_{j}=\beta_{j}(\alpha_{j}),
\end{equation}
here $\alpha_{j}$ , $\beta_{j}$ should satisfy the relation
\begin{equation}\label{lemma4p}
\alpha_{j}^5-25\alpha_{j}^3+30\beta_{j}\alpha_{j}^2+1215\alpha_{j}-\beta_{j}^3-243\beta_{j}=0.
\end{equation}
\end{lemma}
\begin{proof}
First, we assume the solution of eq.(\ref{t31}) has the form
$\widehat{\phi} = X(x)\,Y(y)\,T(t)$ then we have
\begin{equation}\label{CKPeigenfun}
\begin{cases}
  \frac{Y_y}{Y} &= \frac{X_{xxx}}{X} + 3\,\frac{X_x}{X}, \\
  \frac{T_t}{T} &= \frac{X_{xxxxx}}{X} + 5\,\frac{X_{xxx}}{X} +
  15\,\frac{X_x}{X}.
\end{cases}
\end{equation}
Let
\begin{equation}
    \label{CYT1}
    \frac{Y_y}{Y} = \alpha, \qquad
     \frac{T_t}{T} = \beta,
\end{equation}
where $\alpha$ and $\beta$ are constants, eq.(\ref{CKPeigenfun}) become
\begin{equation}\label{lemma4pf}
    \begin{cases}
        X_{xxx} &= \alpha \, X - 3 \, X_x, \\
        X_{xxxxx} &= \beta \, X - 15 \, X_x - 5 \, X_{xxx},
    \end{cases}
\end{equation}
which can be further reduced to
\begin{equation}\label{lemma4pf1}
  \begin{cases}
    9 \, X_{xx} - ( \alpha + \beta ) \, X_x + \alpha^2 \, X &=0, \\
    \alpha \, X_{xx} + 9 \, X_x + (2 \, \alpha - \beta) \, X &=0.
  \end{cases}
\end{equation}
Combining the two equations in eq.(\ref{lemma4pf1}) together, we have
\begin{equation}\label{lemma4pf2}
  (\alpha^2 + \alpha \, \beta + 81) \, X_x = (\alpha^3 - 18\,\alpha
  + 9\,\beta)\,X.
\end{equation}
The solution of eq.(\ref{lemma4pf2})
\begin{equation}\label{lemma4pf3}
  X(x) =
  c_1\,e^{\frac{\alpha^{3}-18\alpha+9\beta}{\alpha^{2}+\alpha\beta+81}x}.
\end{equation}
By substituting eq.(\ref{lemma4pf3}) back into eq.(\ref{lemma4pf}), we
have
\begin{equation}\label{constrain2}
  \alpha^5-25\alpha^3+30\beta\alpha^2+1215\alpha-\beta^3-243\beta=0,
\end{equation}
 that means if $\alpha$ and $\beta$ satisfy eq.(\ref{constrain2}),
then eq.(\ref{lemma4pf3}) is the solution of eq.(\ref{lemma4pf}).
From eq.(\ref{CYT1}), we have
\begin{equation*}
  Y(y)  =  c_2\, e^{\alpha y}, \qquad
  T(t)  =  c_3\, e^{\beta t},
\end{equation*}
together with eq.(\ref{lemma4pf3}) we have
\begin{equation}
  \widehat{\phi} = k \, e^{\frac{\alpha^{3}-18\alpha+9\beta}{\alpha^{2}+\alpha\beta+81}x+\alpha y+\beta t},
\end{equation}
where $k=c_1\,c_2\,c_3$. Using the linear superposition as we did in
lemma \ref{lemma2}, we can obtain
\begin{equation}
  \phi_{i}^{(0)} = \sum_{j=1}^{n} k_j \, \widehat{\phi_j}  =\sum_{j=1}^{n} k_j \,
e^{\frac{\alpha_{j}^{3}-18\alpha_{j}+9\beta_{j}}{\alpha_{j}^{2}+\alpha_{j}\beta_{j}+81}
x +\alpha_{j} y + \beta_{j} t}.
\end{equation}
\end{proof}

Similar to the previous section about KP equation, we need the
following theorem to assure that the solutions we get are without
singularities.
\begin{theorem}\label{theorem2}
Let eigenfunctions $\phi_{m}^{(0)}$ take the form as in lemma \ref{lemma4}
\begin{equation}\label{theorem2eq1}
     \phi_{m}^{(0)}=\sum_{i=1}^{n}
     k_{m,i}\,e^{a_{m,i}x+\alpha_{m,i}y+\beta_{m,i}t},
\end{equation}
where $m=1,2$, if $k_{m,i}>0$, $a_{1,i}<a_{2,j}$, then ${\rm
IW}_{2,2}(\phi_{2}^{(0)},\phi_{1}^{(0)};\phi_{1}^{(0)},\phi_{2}^{(0)})<0$.
The solution of the (2+1)-dimensional {\rm KK} equation can be
written as
\begin{equation}
  \label{ckpu2}
  u_2^{(2+2)} = 1 + \left( \log {\rm IW}_{2,2} \right)_{xx}
\end{equation}
\end{theorem}
\begin{proof}
We rewrite $\phi_1^{(0)}$ and $\phi_2^{(0)}$ in
eq.(\ref{theorem2eq1}) as
\begin{equation*}
\begin{cases}
\phi_1^{(0)} & = \sum_{i=1}^{n} R_{i} e^{a_{i} x}, \\
\phi_2^{(0)} & = \sum_{i=1}^{n} S_{i} e^{b_{i} x}.
\end{cases}
\end{equation*}
Here the values of $R_{i}$ and $S_{i}$ are greater than zero. Then we have
\begin{eqnarray}\label{ckpp1}
\int (\phi_1^{(0)})^2 &= \sum_{i,j=1}^{n} R_{i} R_{j} \frac{e^{(a_i
+ a_j)x}}{a_i + a_j},\\\label{ckpp2}
 \int (\phi_2^{(0)})^2 &=
\sum_{i,j=1}^{n} S_{i} S_{j} \frac{e^{(b_i + b_j)x}}{b_i + b_j},\\
\label{ckpp3}\int \phi_1^{(0)} \phi_2^{(0)} &=\sum_{i,j=1}^{n} R_{i}
S_{j} \frac{e^{(a_i + b_j)x}}{a_i + b_j}.
\end{eqnarray}
Since $a_{i}<b_{j}$ for $i,j=1\ldots n$, it's easy prove the
following inequality
\begin{equation}\label{ckpp4}
R_{i} R_{j} \frac{e^{(a_i + a_j)x}}{a_i + a_j} S_{k} S_{l}
\frac{e^{(b_k + b_l)x}}{b_k + b_l} > R_{i} S_{k} R_{j} S_{l}
\frac{e^{(a_i + b_k)x}}{a_i + b_k} \frac{e^{(a_j + b_l)x}}{a_j +
b_l},
\end{equation}
where $1 \leq i,j,k,l \leq n$, then
\begin{equation} \label{CKPp}
\begin{vmatrix}
\int \phi_1^{(0)} \phi_2^{(0)} & \int (\phi_1^{(0)})^2 \\
\int (\phi_2^{(0)})^2 & \int \phi_1^{(0)} \phi_2^{(0)}
\end{vmatrix}
=(\int \phi_1^{(0)} \phi_2^{(0)})^2 - \int (\phi_1^{(0)})^2 \int
(\phi_2^{(0)})^2 < 0.
\end{equation}
can be directly verified by using eq.(\ref{ckpp1}),
eq.(\ref{ckpp2}), eq.(\ref{ckpp3}). Eq.(\ref{ckpu2}) can be obtained
by eq.(\ref{CKPgtnplusn}) and $u_2^{(0)}=1$.
\end{proof}

\begin{remark}
  For $T_{(1+1)} = T_I\, T_D$, with the generating function $\phi_1^{(0)}$ as in
  eq.(\ref{theorem2eq1}), it's easy to show that
  \begin{equation}\label{remark1eq1}
  \tau^{(1+1)} = (\int (\phi_1^{(0)})^2) \, \tau^{(0)}
  \end{equation}
  is positive. The corresponding new solution of the (2+1)-dimensional {\rm KK} equation can be represented as
  \begin{equation}
    u_2^{(1+1)} = 1 + (\log \int (\phi_1^{(0)})^2 )_{xx}
  \end{equation}
\end{remark}

Here we give the single-soliton solution of the (2+1)-dimensional KK equation from
the generating function
\begin{equation}
\phi_1^{(0)} = e^{\xi_1} + e^{\xi_2},
\end{equation}
where $\xi_i = \frac{\alpha_i^3 - 18 \alpha_i + 9
\beta_i}{\alpha_i^2 + \alpha_i \beta_i + 81} \, x + \alpha_i \, y +
\beta_i \, t$, the solution is
\begin{equation}\label{CKPsinglesoliton}
u_2^{(1+1)}= 1 + \frac{(a_1-a_2)^2}{a_1+a_2}
\frac{(\frac{e^{\frac{\xi_1-\xi_2}{2}}}{a_1}+\frac{e^{\frac{\xi_2-\xi_1}{2}}}{a_2})
\, ( e^{\frac{\xi_1-\xi_2}{2}} +
e^{\frac{\xi_2-\xi_1}{2}})}{(\frac{e^{\xi_1-\xi_2}}{2a_1} +
\frac{e^{\xi_2-\xi_1}}{2a_2} + \frac{2}{a_1+a_2} )^2},
\end{equation}
where $a_i = \frac{\alpha_i^3 - 18 \alpha_i + 9 \beta_i}{\alpha_i^2
+ \alpha_i \beta_i + 81}$. The solution $(u_2^{(1+1)})^{\prime}$
generated from zero seed have the form
\begin{equation}\label{CKPsinglesoliton0}
(u_2^{(1+1)})^{'}= \frac{(a_1^{'}-a_2^{'})^2}{a_1^{'}+a_2^{'}}
\frac{(\frac{e^{\frac{\xi_1^{'}-\xi_2^{'}}{2}}}{a_1^{'}}+\frac{e^{\frac{\xi_2^{'}-\xi_1^{'}}{2}}}{a_2^{'}})
\, ( e^{\frac{\xi_1^{'}-\xi_2^{'}}{2}} +
e^{\frac{\xi_2^{'}-\xi_1^{'}}{2}})}{(\frac{e^{\xi_1^{'}-\xi_2^{'}}}{2
a_1^{'}} + \frac{e^{\xi_2^{'}-\xi_1^{'}}}{2 a_2^{'}} +
\frac{2}{a_1^{'}+a_2^{'}} )^2},
\end{equation}
where $\xi_i^{'} = \frac{(\alpha_i^{'})^2}{\beta_i^{'}} \, x +
\alpha_i^{'} \, y + \beta_i^{'} \, t$,
$a_i^{'}=\frac{(\alpha_i^{'})^2}{\beta_i^{'}}$ and
$(\alpha_i^{'})^5=(\beta_i^{'})^3$. The differences between
$u_2^{(1+1)}$ and $(u_2^{(1+1)})^{'}$ under the same value of
parameters are showed in figure 3.
By taking
\begin{eqnarray}\nonumber
  \phi_1^{(0)} &=& {e^{ 0.0001999999974\,x+ 0.0006\,y+ 0.003\,t}}+{e^{
 0.0006666665679\,x + 0.002\,y+ 0.01\,t}}\\&&+{e^{ 0.003333320988
\,x+ 0.01\,y+ 0.05\,t}}+{e^{ 0.006666567904\,x+ 0.02 \,y+ 0.1\,t}},
\\\nonumber
  \phi_2^{(0)}&=& {e^{ 1.218304787\,x+ 5.463203409\,y+30\,t}}+{e^{ 0.4917724251\,x+
 1.594247576\,y+8\,t}}\\&&+{e^{ 0.6835764081\,x+ 2.370148557\,y+12\,t}}+{e
^{ 0.970831384\,x+ 3.827515914\,y+20\,t}}.
\end{eqnarray}
in eq.(\ref{ckpu2}), we can obtain solution of the
($2+1$)-dimensional KK equation which is plotted in figure 4.

\section{Gauge transformation for (2+1)-dimensional SK equation}
 The procedure of this section is mostly the same as the previous section
except that the transformed Lax operator $L^{(1)}$ by one-step gauge
transformation should satisfy $(L^{(1)})^{*} = -
\partial L^{(1)} \partial^{-1}$, so we need lemma \ref{lemma5} about
gauge transformation for BKP hierarchy.
\begin{lemma}[\cite{Jhep}]
    \label{lemma5}
    \begin{enumerate}
    \item The appropriate gauge transformation $T_{n+k}$ is given by $n=k$ and generating functions $\psi_{i}^{(0)}=\phi_{i,x}^{(0)}$ for $i=1,2,\ldots,n$.
    \item The $\tau$-function $\tau_{{\rm BKP}}^{(n+n)}$ of the {\rm BKP} hierarchy has the form
      \begin{eqnarray}\nonumber
        \tau_{\text{\rm BKP}}^{(n+n)} &=& {\rm IW}_{n,n}(\phi_{n,x}^{(0)},\phi_{n-1,x}^{(0)},\ldots,\phi_{1,x}^{(0)};\phi_{1}^{(0)},\phi_{2}^{(0)},\ldots,\phi_{n}^{(0)})\cdot
        \tau_{\text{\rm BKP}}^{(0)}\\
        &=&
        \begin{vmatrix}
          \int \phi_{n,x}^{(0)}\cdot\phi_{1}^{(0)} & \cdots & \int \phi_{n,x}^{(0)}\cdot\phi_{n}^{(0)} \\
          \vdots & \cdots & \vdots \\
          \int \phi_{1,x}^{(0)}\cdot\phi_{1}^{(0)} & \cdots & \int \phi_{1,x}^{(0)}\cdot\phi_{n}^{(0)}
        \end{vmatrix}
        \cdot
        \tau_{\text{\rm BKP}}^{(0)}.
      \end{eqnarray}
    \end{enumerate}
    and we have
    \begin{equation}\label{BKPgtnplusn}
    u_{2}^{(n+n)} =u_{2}^{(0)} +
    (\log{{\rm IW}_{n,n}})_{xx}.
    \end{equation}
  \end{lemma}
With this theorem, we can write down the solutions of the
(2+1)-dimensional SK equation explicitly after successive
application of gauge transformations. We take the initial Lax
operator $L^{(0)}$ of the BKP hierarchy  as
  \begin{equation*}
    L^{(0)}=\partial+\partial^{-1}+\partial^{-3}+\partial^{-5}+\cdots.
  \end{equation*}
The corresponding eigenfunction $\phi_{i}^{(0)}$ and conjugate
eigenfunction $\psi_{i}^{(0)}=\phi_{i,x}^{(0)}$  are given by lemma
\ref{lemma4} and lemma \ref{lemma5}, i.e.
\begin{eqnarray}
\label{skphi} \phi_{i}^{(0)} &=&\sum_{j=1}^{n} k_j \,
e^{\frac{\alpha_{j}^{3}-18\alpha_{j}+9\beta_{j}}{\alpha_{j}^{2}+\alpha_{j}\beta_{j}+81}
x +\alpha_{j} y + \beta_{j} t},
\\ \psi_{i}^{(0)} &=&\sum_{j=1}^{n} k_j \,
\frac{\alpha_{j}^{3}-18\alpha_{j}+9\beta_{j}}{\alpha_{j}^{2}+\alpha_{j}\beta_{j}+81}
\,
e^{\frac{\alpha_{j}^{3}-18\alpha_{j}+9\beta_{j}}{\alpha_{j}^{2}+\alpha_{j}\beta_{j}+81}
x +\alpha_{j} y + \beta_{j} t},\qquad
\beta_{j}=\beta_{j}(\alpha_{j}).
\end{eqnarray}
Similar as section two and section three, we need the following
theorem to assure that the new $\tau$-function we get after gauge
transformations will not vanish at any point.
\begin{theorem}\label{theorem3}
Let eigenfunction $\phi_{m}^{(0)}$ take the form as in eq.(\ref{skphi})\\
    $\sum_{i=1}^{n} k_{m,i}\,e^{a_{m,i}x+\alpha_{m,i}y+\beta_{m,i}t}$,  $m=1,2$,  if    $ \,0<3\cdot a_{1,i}< a_{2,j}$,\\
    then we have ${\rm IW}_{2,2}(\phi_{2,x}^{(0)},\phi_{1,x}^{(0)};\phi_{1}^{(0)},\phi_{2}^{(0)})<0$.
    The solution can be written as
\begin{equation}
  \label{bkpu2}
  u_2^{(2+2)} = 1 + \left( \log {\rm IW}_{2,2} \right)_{xx}.
\end{equation}
\end{theorem}
\begin{proof}
$\phi_1^{(0)}$ and $\phi_2^{(0)}$ can be rewritten as
\begin{equation*}
\begin{cases}
\phi_1^{(0)} & = \sum_{i=1}^{n} R_{i} e^{a_{i} x}, \\
\phi_2^{(0)} & = \sum_{i=1}^{n} S_{i} e^{b_{i} x},
\end{cases}
\end{equation*}
where the value of $R_{i}$ and $S_{i}$ are greater than zero, then
we have
\begin{eqnarray}\label{bkpp1}
\frac{(\phi_1^{(0)})^2}{2} &=& \frac{1}{2} \, \sum_{i,j=1}^{n} R_{i}
R_{j} e^{(a_i + a_j)x},\\\label{bkpp2} \frac{(\phi_2^{(0)})^2}{2}
&=& \frac{1}{2} \, \sum_{i,j=1}^{n} S_{i} S_{j} e^{(b_i +
b_j)x},\\\label{bkpp3} \int \phi_{1,x}^{(0)} \phi_2^{(0)} &=&
\sum_{i,j=1}^{n} R_{i} S_{j}
\frac{a_i}{a_i+b_j}e^{(a_i+b_j)x},\\\label{bkpp4} \int
\phi_{2,x}^{(0)} \phi_1^{(0)} &=& \sum_{i,j=1}^{n} R_{j} S_{i}
\frac{b_i}{a_j+b_i}e^{(a_j+b_i)x}.
\end{eqnarray}
The following inequality
\begin{equation*}
(a_i + b_k) \, (a_j + b_l) > 4 \, a_i \, b_l,
\end{equation*}
is trivial if we use $0< 3 \cdot a_{1,i} < a_{2,j}$ which means $0<
3\cdot a_i < b_k$, together with eq.(\ref{bkpp1}), eq.(\ref{bkpp2}),
eq.(\ref{bkpp3}) and eq.(\ref{bkpp4}), we can prove
\begin{equation}
\label{BKPp}
\begin{vmatrix}
\int \phi_1^{(0)} \phi_{2,x}^{(0)} &  \frac{(\phi_2^{(0)})^2}{2} \\
\frac{(\phi_1^{(0)})^2}{2} & \int \phi_{1,x}^{(0)} \phi_2^{(0)}
\end{vmatrix}
=(\int \phi_{1,x}^{(0)} \phi_2^{(0)})(\int \phi_{2,x}^{(0)}
\phi_1^{(0)}) - \frac{(\phi_1^{(0)})^2 (\phi_2^{(0)})^2}{4} < 0,
\end{equation}
by a direct calculation. Eq.(\ref{bkpu2}) can be obtained by
eq.(\ref{BKPgtnplusn}) and $u_2^{(0)}=1$.
\end{proof}
\begin{remark}
  For $T_{1+1} = T_I\, T_D$, with the generating function $\phi_1^{(0)}$ as in
  eq.(\ref{skphi}), it's easy to show that
  \begin{equation}\label{remark1eq1}
  \tau^{(1+1)} = \frac{(\phi_1^{(0)})^2}{2} \, \tau^{(0)}
  \end{equation}
  is positive. The corresponding new solution of the (2+1)-dimensional {\rm SK} equation can be represented as
  \begin{equation}
    u_2^{(1+1)} = 1 + (\log ( \frac{(\phi_1^{(0)})^2}{2} ))_{xx}
  \end{equation}
\end{remark}

To obtain a single-soliton solution of the (2+1)-dimensional SK
equation, we start from a generating function
\begin{equation}
  \phi_1^{(0)} = e^{\xi} + e^{-\xi},
\end{equation}
and the solution is
\begin{equation}\label{BKPsinglesoliton}
  u_2^{(1+1)} = 1 + 2\,a^2\,{\rm sech}^2(\xi),
\end{equation}
here $\xi = \frac{\alpha^3 - 18 \alpha + 9 \beta}{\alpha^2 + \alpha
\beta + 81} \, x + \alpha \, y + \beta \, t$ and $a =\frac{\alpha^3
- 18 \alpha + 9 \beta}{\alpha^2 + \alpha \beta + 81}$. A solution
generated from zero seed is
\begin{equation}
  \label{BKPsinglesoliton0}
  (u_2^{(1+1)})^{'} =  2\,(a^{'})^2\,{\rm sech}^2(\xi^{'}),
\end{equation}
in which $\xi^{'} = \frac{(\alpha^{'})^2}{\beta^{'}} \, x +
\alpha^{'} \, y + \beta^{'} \, t$, $(\alpha^{'})^5=(\beta^{'})^3$
and $a^{'}=\frac{(\alpha^{'})^2}{\beta^{'}}$. The differences
between $u_2^{(1+1)}$ and $(u_2^{(1+1)})^{'}$ are showed in figure
5. In figure 6, we plot the solution of the ($2+1$)-dimensional SK
equation by taking
\begin{eqnarray}\nonumber
  \phi_1^{(0)} & = & {e^{ 0.009999666694\,x+ 0.02999999998\,y+ 0.15\,t}}+{e^{ 0.01333254332
     \,x+ 0.03999999992\,y+ 0.2\,t}} \\ &&+{e^{ 0.006666567904\,x+
     0.02 \,y+ 0.1\,t}},\\\nonumber
  \phi_2^{(0)} & = & {e^{ 0.5924749002\,x+ 1.985399095\,y+10\,t}}+{e^{ 0.06656825084\,x+
 0.1999997386\,y+t}}\\&&+{e^{ 1.218304787\,x+
 5.463203409\,y+30\,t}},
\end{eqnarray}
in eq.(\ref{bkpu2}).
\begin{corollary} \label{corskkktrans}
  For the (2+1)-dimensional {\rm KK} equation and (2+1)-dimensional {\rm
  SK}
  equation, there exist a common Galilean type transformation
  between $(u_2^{(1+1)})^{\prime}$ (generated from zero seed) and
  $u_2^{(1+1)}$ (generated from non-zero seed), i.e.
  \begin{equation}\label{BCKPtrans}
    u^{\prime}_2(x,y,t) \longmapsto  u_2(x,y,t)= 1 + u_2^{'} (x+3y+15t, y+5t, t ).
  \end{equation}
\end{corollary}

\section{Conclusions and Discussions}
By now we have obtained new solutions $u_2^{(n)}$ in theorem
\ref{theorem1} for KP equation, $u_2^{(2+2)}$ in theorem
\ref{theorem2} for (2+1)-dimensional KK equation and $u_2^{(2+2)}$
in theorem \ref{theorem3} for (2+1)-dimensional SK equation by using
the the gauge transformations of the KP hierarchy, CKP hierarchy and
BKP hierarchy respectively. The corresponding generating functions
of the gauge transformations previously mentioned are explicitly
expressed in lemma \ref{lemma2} and lemma \ref{lemma4}. For these
three equations, the single-soliton $u^{(1)}_2$(or $u_2^{(1+1)}$)
generated from non-zero seeds and $(u_2^{(1)})^{'}$(or
$(u_2^{(1+1)})^{'}$  ) generated from zero seed are constructed. The
main differences between the  $u_2$ and  $(u_2)^{'}$ are height and
locations of the peak line under the same value of parameters, which
are demonstrated visibly in figures 1, 2 and 3. We also found a
Galilean type transformation in eq.(\ref{kptrans}) between
$(u_2^{(1)})^{'}$ and $ u_2^{(1)}$ for the KP equation, and another
one in eq.(\ref{BCKPtrans}) between $(u_2^{(1+1)})^{'}$ and $
u_2^{(1+1)}$ for the (2+1)-dimensional KK and SK equations. To
guarantee the new solutions $u_2$ generated by gauge transformations
is smooth, in other words, the transformed $\tau$-function doesn't
vanish at any point, we only consider the ${\rm W}_n$ in theorem
\ref{theorem1} and ${\rm IW}_{2,2}$ in theorem \ref{theorem2} and
theorem \ref{theorem3}.

 The corollary 1 and corollary 2 show that we can establish a one-parameter
transformation group (specifically, Galilean type transformation) of
the solutions of these three equations by setting the seeds
$u_2^{(0)}=\epsilon$(arbitrary constant) instead of $u_2^{(0)}=1$.
The advantage of this new method to find one-parameter group is to
avoid solving the characteristic line equation, which is not easy to
solve, as usual approach of Lie point transformation. We will try to
do this in the future. On the other hand, if we can  choose some
more complicated initial Lax operator $L^{(0)}$ in which
\{$u^{(0)}_i$\} are not constants and we are able to solve the
corresponding generating functions, then we can get some other new
solutions. Of course, the calculation is much tedious although the
idea is straightforward. The present work is the first step to this
difficult purpose.

\vskip 1cm

{\bf Acknowledgement}\\
{\small This work is supported partly by the NSFC grant of China
under  No.10671187. We thank Professor Li Yishen(USTC, China) for
many valuable suggestions on this paper. }


\newpage

\begin{figure}\label{figkpcompare}
    \includegraphics[width=3in,totalheight=2in]{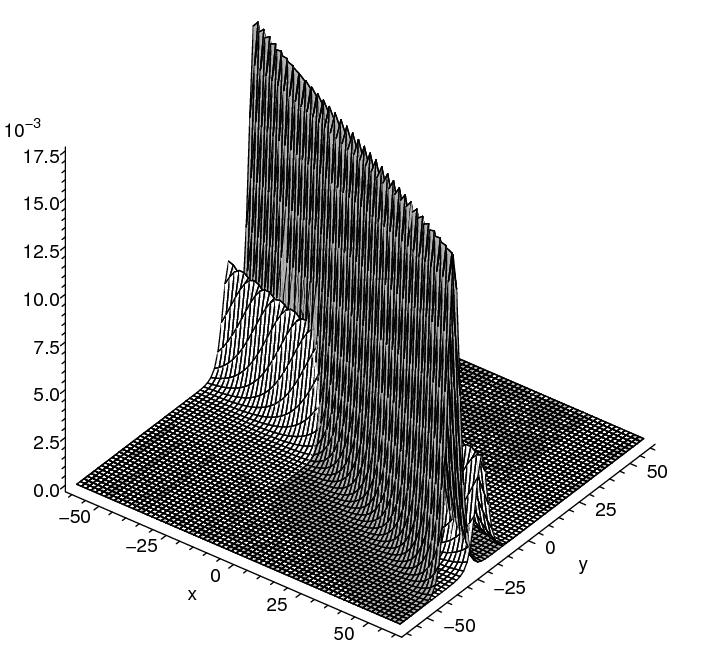}
    \caption{Single-soliton solutions at $t=1$ of the KP equation.
      The lower one is $(u_2^{(1)})^{'}$ with $k^{'}=1$, $\alpha_1^{'}=2.7225$ and
    $\alpha_2^{'}=3.24$; the higher one is $(u_2^{(1)}-1)$ with parameters $k=1$, $\alpha_1=2.7225$ and $\alpha_2=3.24$.}
  \end{figure}

\begin{figure}\label{figtwostepkp}
    \includegraphics[width=3.5in,totalheight=2.3in]{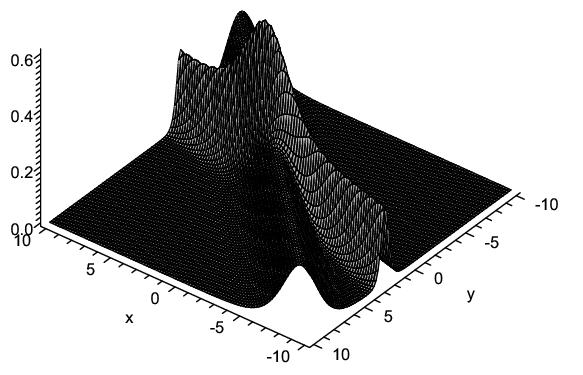}
    \caption{Two-soliton solution at $t=0$ of the KP equation.}
  \end{figure}

\begin{figure}\label{fig2dkkcompare}
    \includegraphics[width=3in,totalheight=2in]{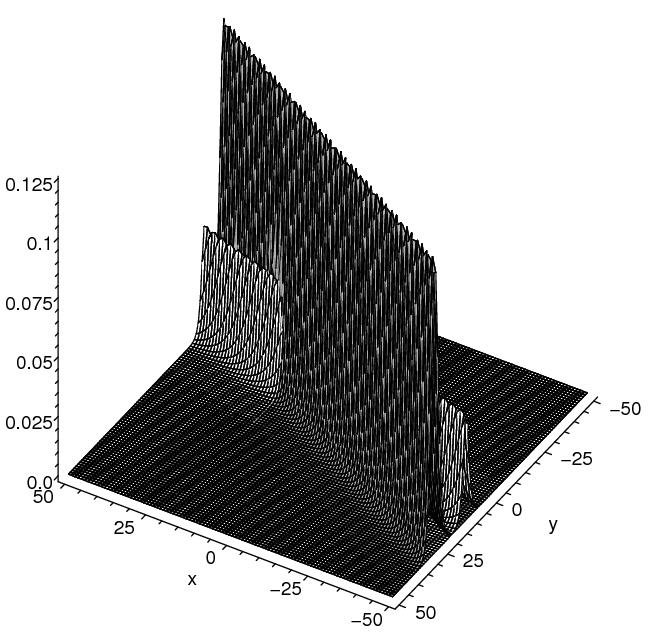}
    \caption{Single-soliton solutions at $t=1$ of the
    (2+1)-dimensional KK equation. The higher one is $(u_2^{(1+1)})^{'}$ with $\alpha_1^{'}=0.970299$ and
    $\alpha_2^{'}=0.075$; the lower one is $(u_2^{(1+1)}-1)$ with parameters $\alpha_1=0.970299$ and $\alpha_2=0.075$.}
  \end{figure}

\begin{figure}\label{figtwostep2dkk}
    \includegraphics[width=5in,totalheight=3.3in]{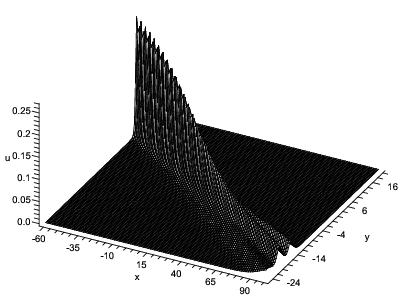}
    \caption{Solution at $t=0$ of the
    (2+1)-dimensional KK equation.}
  \end{figure}

\begin{figure}\label{fig2dskcompare}
    \includegraphics[width=3in,totalheight=2in]{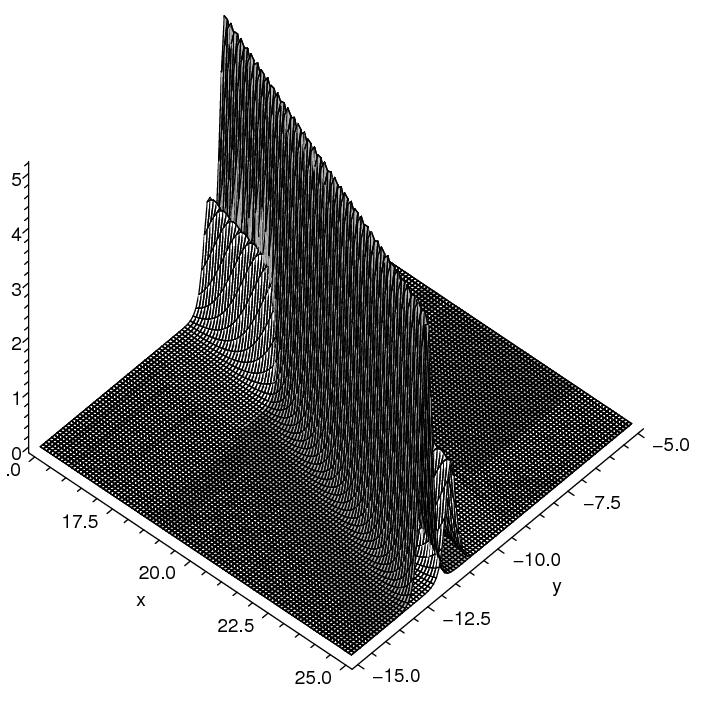}
    \caption{Single-soliton solutions at $t=1$ of (2+1)-dimensional SK equation. The higher one is $(u_2^{(1+1)})^{'}$ with $\alpha^{'}=4.096$
,the lower one is $(u_2^{(1+1)}-1)$ with parameters $\alpha=4.096$}.
  \end{figure}

\begin{figure}\label{figtwostep2dsk}
    \includegraphics[width=5in,totalheight=3.3in]{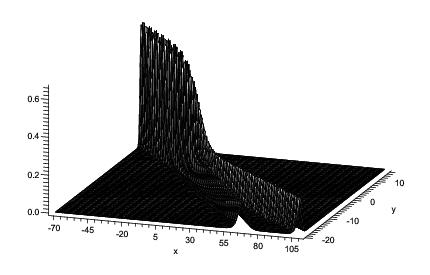}
    \caption{Solution at $t=0$ of (2+1)-dimensional SK equation.}
  \end{figure}


\begin{thebibliography}{200}
\bibitem{ablowitz1}J. Villarroel, M.J. Ablowitz, On the initial value problem for the KPII
equation with data that does not decay along a line, Nonlinearity 17 (2004), 1843-1866.
\end{thebibliography}
\end{document}